\documentclass{aastex63}

\newcommand{\sdo}{\textit{SDO}}
\newcommand{\rhessi}{\textit{RHESSI}}
\newcommand{\hinode}{\textit{Hinode}}
\newcommand{\iris}{\textit{IRIS}}

\begin{document}
	
\title{Observations of magnetic reconnection and particle acceleration locations in solar coronal jets}

\author{Yixian Zhang}
\affiliation{School of Physics and Astronomy, University of Minnesota, 116 Church St. SE, Minneapolis, MN 55455,USA}
\author{Sophie Musset}
\affiliation{European Space Agency (ESA), European Space Research and Technology Centre (ESTEC), Keplerlaan 1, 2201 AZ, Noordwijk, The Netherlands}
\author{Lindsay Glesener}
\affiliation{School of Physics and Astronomy, University of Minnesota, 116 Church St. SE, Minneapolis, MN 55455,USA}
\author{Navdeep Panesar}
\affiliation{Lockheed Martin Solar Astrophysics Laboratory, 3251 Hanover Street Building 252, Palo Alto, CA 94304, USA}
\affiliation{Bay Area Environmental Research Institute, NASA Research Park, Moffett Field, CA 94035, USA}
\author{Gregory Fleishman}
\affiliation{Center For Solar-Terrestrial Research, New Jersey Institute of Technology, Newark, NJ 07102, USA}

\begin{abstract}		
	We present a multi-wavelength analysis of two flare-related jets on November 13, 2014, using data from {\sdo}/AIA, {\rhessi}, {\hinode}/XRT, and {\iris}. Unlike most coronal jets where hard X-ray (HXR) emissions are usually observed near the jet base, in these events HXR emissions are found at several locations, including in the corona. We carry out the first differential emission measure (DEM) analysis that combines both AIA (and XRT when available) bandpass filter data and {\rhessi} HXR measurements for coronal jets, and obtain self-consistent results across a wide temperature range and into non-thermal energies. In both events, hot plasma first appeared at the jet base, but as the base plasma gradually cooled, hot plasma also appeared near the jet top. Moreover, non-thermal electrons, while only mildly energetic, are found in multiple HXR locations and contain a large amount of total energy. Particularly, the energetic electrons that produced the HXR sources at the jet top were accelerated near the top location, rather than traveling from a reconnection site at the jet base. This means that there was more than one particle acceleration site in each event. Jet velocities are consistent with previous studies, including upward and downward velocities around $\sim$200 km/s and $\sim$100 km/s respectively, and fast outflows of 400-700 km/s. We also examine the energy partition in the later event, and find that the non-thermal energy in accelerated electrons is most significant compared to other energy forms considered. We discuss the interpretations and provide constraints on mechanisms for coronal jet formation.
	
\end{abstract}

\section{Introduction} \label{sec:intro}
Solar coronal jets are collimated plasma ejections that occur in the solar corona and they offer ways for plasma and particles to enter interplanetary space. They are transient (tens of minutes) but ubiquitous, with a typical height of $\sim5\times10^4$ km and a typical width of $\sim8\times10^3$ km \citep[e.g.][]{2007PASJ...59S.771S}. X-ray emissions from coronal jets were first observed by the Soft X-ray Telescope (SXT) onboard \textit{Yohkoh} in the early 1990s \citep{1992PASJ...44L.173S}. Since then, coronal jets have been studied in various aspects including morphology, dynamics, driving mechanisms, and more \citep[see, e.g.,][]{2016SSRv..201....1R, 2021RSPSA.47700217S}. Jets or jet-like events have also been observed in other wavelengths such as extreme ultraviolet (EUV), ultraviolet (UV), and H alpha (those studied in H alpha are historically known as ``surges'') \citep[e.g.][]{1997Natur.386..811I, 1999ApJ...513L..75C,2009SoPh..259...87N}. These wavebands cover a wide range of plasma temperatures (from chromospheric to coronal), and one important feature of jets is the presence of both hot and cool components in many events \citep{2010ApJ...720..757M, 2013ApJ...769..134M}.

Current models generally suggest that jets are formed by magnetic reconnection between open and closed magnetic field lines; however, the detailed triggering process for such magnetic reconnection is still not fully clear. In the emerging flux model, jets are generated through interchange reconnection when field lines of the newly emerging magnetic flux reach those of the pre-existing open field \citep{1992PASJ...44L.173S}. As shown in the 2D simulation by \citet{1995Natur.375...42Y, 1996PASJ...48..353Y}, this model could successfully produce a hot jet and an adjacent cool jet (or surge) simultaneously. The embedded-bipole model developed by \citet[][etc.]{2015A&A...573A.130P, 2016A&A...596A..36P} considers a 3D fan-spine topology where magnetic reconnection occurs around the 3D null point. In their simulation, straight jets are generated through slow reconnection at the current sheet and driven by magnetic tension, while helical jets are generated through explosive magnetic reconnection triggered by a kink-like instability and driven by a rapid untwisting process of magnetic field lines. Recently, a few studies have reported small-scale filament structures (known as ``minifilaments'') at the base of some coronal jets, leading to the minifilament eruption model \citep{2015Natur.523..437S, 2016ApJ...821..100S, 2016ApJ...832L...7P, 2017ApJ...844..131P, 2018ApJ...853..189P, 2017Natur.544..452W, 2018ApJ...852...98W, 2018ApJ...859....3M, 2019ApJ...882...16M}. This model suggests that jets are generated through miniature filament eruptions similar to those that drive larger eruptive events such as coronal mass ejections (CMEs). In addition to the external/interchange magnetic reconnection, this process also involves internal magnetic reconnection inside the filament-carrying field, and the jet bright point (JBP, which corresponds to the solar flare arcade in the larger-scale case) appears underneath the erupting minifilament. Many recent observations have shown that the triggers for these minifilaments eruptions are usually magnetic flux cancellation \citep{2011ApJ...738L..20H, 2012A&A...548A..62H, 2014ApJ...783...11A,  2016ApJ...832L...7P, 2017ApJ...844..131P, 2018ApJ...853..189P, 2019ApJ...882...16M, 2021ApJ...909..133M}.

HXR observations can also provide helpful insights into jet formation mechanisms by constraining energetic electron populations within coronal jets. \citet{2011ApJ...742...82K} investigated HXR emissions for 16 flare-related energetic electron events and found that 7 of them showed three distinct HXR footpoints, which was consistent with the interchange reconnection geometry. (In the remaining events, the fact that they showed less then three sources was likely due to instrument limitations.) Also in that study, EUV jets were found in all 6 events that had EUV data coverage. HXR bremsstrahlung emissions could also directly come from coronal jets if there are energetic electrons, but those extended sources are usually much fainter than the footpoint sources and only a few studies \citep{2009A&A...508.1443B, 2012ApJ...754....9G} have reported such observations. More recently, \citet{2018ApJ...867...84G} combined HXR observations with microwave emission, EUV emission, and magnetogram data, performing 3D modeling of electron distributions for a flare-related jet. They obtained direct constraints on energetic electron populations within that event. \citet{2020ApJ...889..183M} carried out a statistical study of 33 flare-related coronal jets using HXR and EUV data, and they observed non-thermal emissions from energetic electrons in 8 of these events. They also studied the relation between jets and the associated flares but found no clear correlations between jet and flare properties. 

In most of the previous studies of coronal jets, hot plasma and HXR emissions were found near the base of the jet (the location of the primary reconnection site) \citep[e.g.][]{2011ApJ...742...82K, 2016A&A...589A..79M, 2020ApJ...889..183M}. However, for two coronal jets on November 13, 2014, HXR thermal emissions were observed near the far end of the jet spire (hereafter the ``top''). In fact, in the second event which had full HXR coverage, HXR emissions were observed at three different locations: the base of the jet, the top of the jet, and a location to the north of the jet. Here we present a multi-wavelength analysis of these two jets using data from the Atmospheric Imaging Assembly (AIA) onboard the \textit{Solar Dynamic Observatory} (\sdo), the \textit{Reuven Ramaty High Energy Solar Spectroscopic Imager} (\rhessi), the X-ray Telescope (XRT) onboard \hinode, and the \textit{Interface Region Imaging Spectrograph} (\iris). We found that all these different HXR sources showed evidence of mildly accelerated electrons, and particle acceleration also happened near the jet top in addition to the site at the jet base. To our knowledge, this is the most thorough HXR study to date of particle acceleration in coronal jets.

The paper is structured as follows: In section \ref{sec:data}, we describe the observations from each instrument. In Section \ref{sec:analysis}, we show results from differential emission measure (DEM) analysis, imaging spectroscopy, and velocity estimation. In Section \ref{sec:discussion}, we calculate the energy budget for one of the jets, discuss the interpretations of observational results, and compare them with jet models. Finally, in Section \ref{sec:summ}, we summarize the key findings of this work.

\section{Observations} \label{sec:data}
\begin{figure}
	\includegraphics[width=0.5\textwidth]{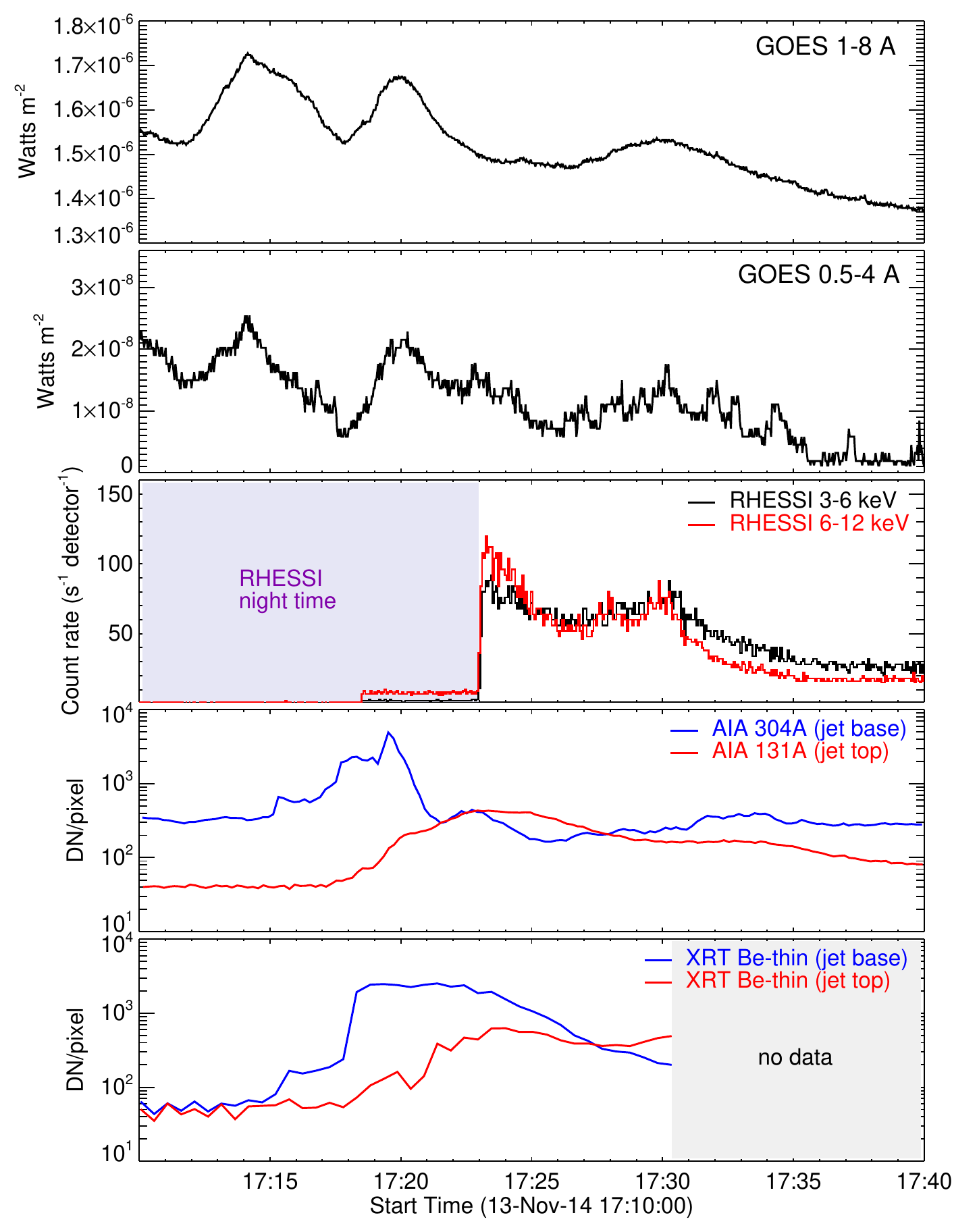}
	\includegraphics[width=0.5\textwidth]{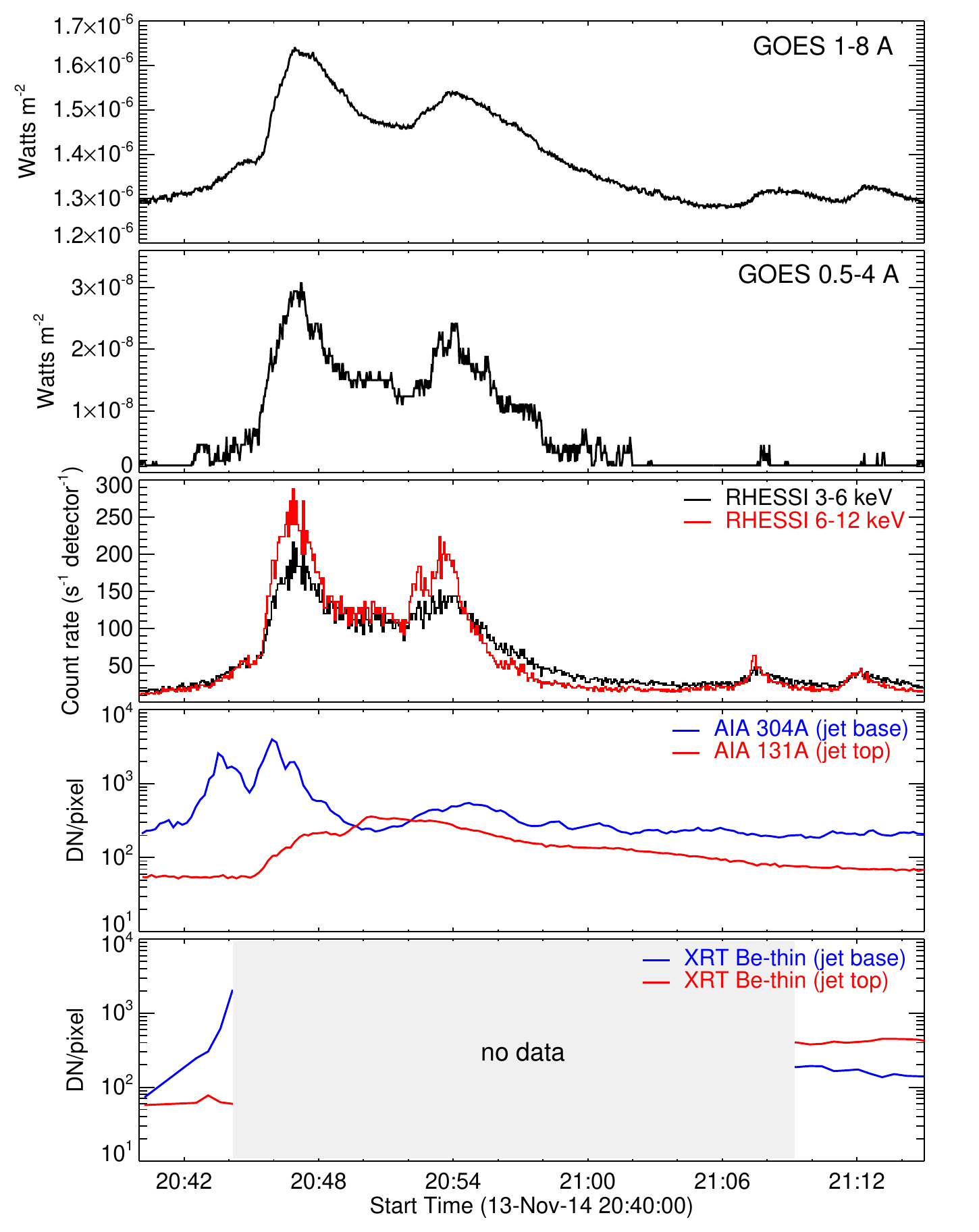}
	\caption{Time profiles of the $\sim$17:20 jet (left) and the $\sim$20:50 jet (right) on November 13, 2014. Top row: GOES light curves in the 1-8 {\AA} channel. Second row: GOES light curves in the 0.5-4 {\AA} channel. Third row: {\rhessi} emission in 3-6 keV (black) and 6-12 keV (red), using detectors 3, 6, 8, and 9. The first event only has partial coverage from {\rhessi} due to spacecraft night. Fourth row: Examples of AIA EUV emissions from the jet base/top. Blue lines show light curves of a 3"$\times$3" box at the base of each jet in the 304 {\AA} channel, and red lines show light curves of a 3"$\times$3" box at the top of each jet in the 131 {\AA} channel (boxes are not shown). Bottom row: XRT measurements of selected regions (3"$\times$3", not shown) at the jet base (blue) and the jet top (red) in the thin-Be filter. Both events only have partial coverage from XRT.}
	\label{fig:t_profile}	
\end{figure}

On November 13, 2014, more than ten recurrent jets were ejected from NOAA Active Region 12209 near the eastern solar limb at different times throughout the day. While most (if not all) of the jets can be identified in one or more AIA channels, only two events, SOL2014-11-13T17:20 and SOL2014-11-13T20:47, were simultaneously observed by AIA and {\rhessi}. We select these two flare-related jets for this study, and we add supporting observations from XRT and {\iris}. The associated flares are GOES class C1.5-1.7 without background subtraction (see top row of Figure \ref{fig:t_profile}), or B2.4-3.7 with background subtraction.

\subsection{AIA data}

The AIA instrument provides full-disk solar images in ten EUV/UV/visible-light channels with a spatial resolution of 1.5 arcsec \citep{2012SoPh..275...17L}. In this work we use data from the seven EUV channels of AIA: 94 {\AA}, 131 {\AA}, 171 {\AA}, 193 {\AA}, 211 {\AA}, 304 {\AA}, and 335 {\AA}, which have a cadence of 12 seconds and cover plasma temperatures from $\sim$0.05 MK up to $\sim$20 MK.

\begin{figure*}
	\centering
	\includegraphics[width=0.98\textwidth]{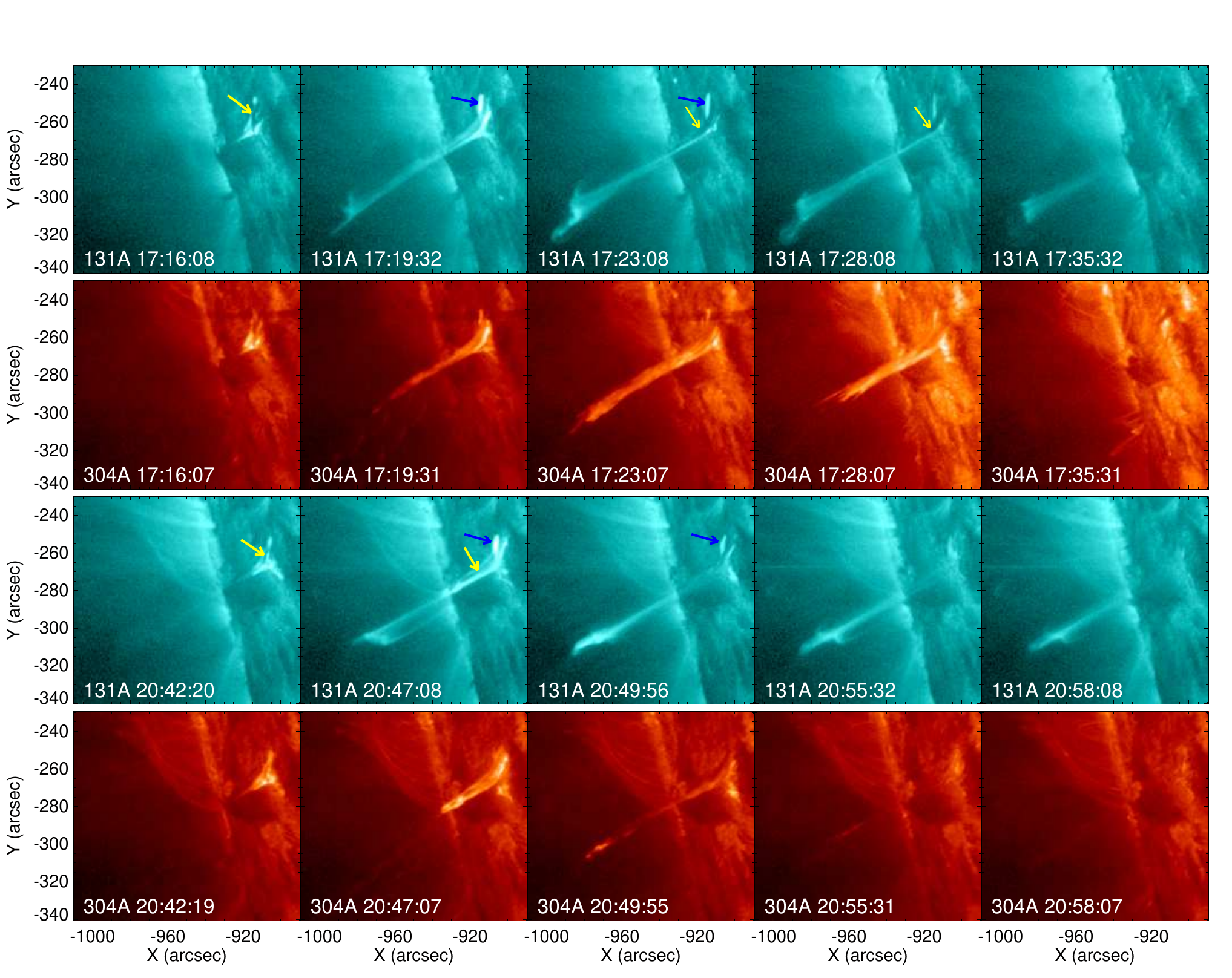}
	\caption{AIA 131 {\AA} and 304 {\AA} images of the two jets at selected times. The top two rows show the evolution of the earlier jet and the bottom two rows show the evolution of the later jet. The yellow arrows point to the minifilament while the blue arrows point to the JBP in each event. Both jets reached their maximum extents at earlier times and lasted longer in the hotter 131 {\AA} channel (sensitive to both $\sim$0.4 MK and $\sim$10 MK temperatures) compared to the cool 304 {\AA} channel (sensitive to chromospheric temperatures around 0.05 MK).}
	\label{fig:aia_im}	
\end{figure*}

Figure \ref{fig:aia_im} shows AIA images of the two jets in the 131 {\AA} and 304 {\AA} channels at selected times. At the beginning of each event, a minifilament (pointed by yellow arrows) was identified in multiple AIA channels at the base of the jet. After the minifilamnet eruption, a JBP (pointed by white arrows) appeared underneath the prior minifilament location.

Interestingly, both events showed slightly different jet evolution in cool and hot AIA channels. In cooler channels including 171 {\AA}, 193 {\AA}, 211 {\AA}, 304 {\AA}, and 335 {\AA}, the first jet started at $\sim$17:15 UT, reached its maximum extent at $\sim$17:23 UT, and lasted about 20 minutes. However, in hot channels that are sensitive to $\gtrsim$10 MK plasma (94 {\AA} and 131 {\AA}, particularly), the jet reached its maximum height within five minutes after the same starting time; then it slightly expanded transversely and gradually faded away in a much longer time. (The 193 {\AA} channel in principle could also measure hot plasma \citep{2012SoPh..275...17L}, but its response was dominated by temperatures below 10 MK and thus looked like a cool channel.) Similar behavior was observed in the later jet, which started at $\sim$20:40 UT and reached its maximum extent at $\sim$20:47 UT in the hotter 94 {\AA} and 131 {\AA} channels or at $\sim$20:50 UT in the rest of the channels. The jet had already disappeared in those cooler channels before 20:58 UT, but it was visible in 94 {\AA} and 131 {\AA} for more than an hour. 

\subsection{RHESSI data} \label{sec:rhessi}

{\rhessi} was a solar-dedicated HXR observatory launched in 2002 and decommissioned in 2018. It consisted of nine rotating modulation collimators, each placed in front of a cooled germanium detector, and used indirect Fourier imaging techniques. {\rhessi} measured both images and spectra over the full sun in the energy range of 3 keV - 17 MeV and had good spatial and energy resolutions especially for lower energies (2.3 arcsec and $\sim$1 keV, respectively) \citep{2002SoPh..210....3L}. 

\begin{figure*}
	\centering
	\includegraphics[width=0.99\textwidth]{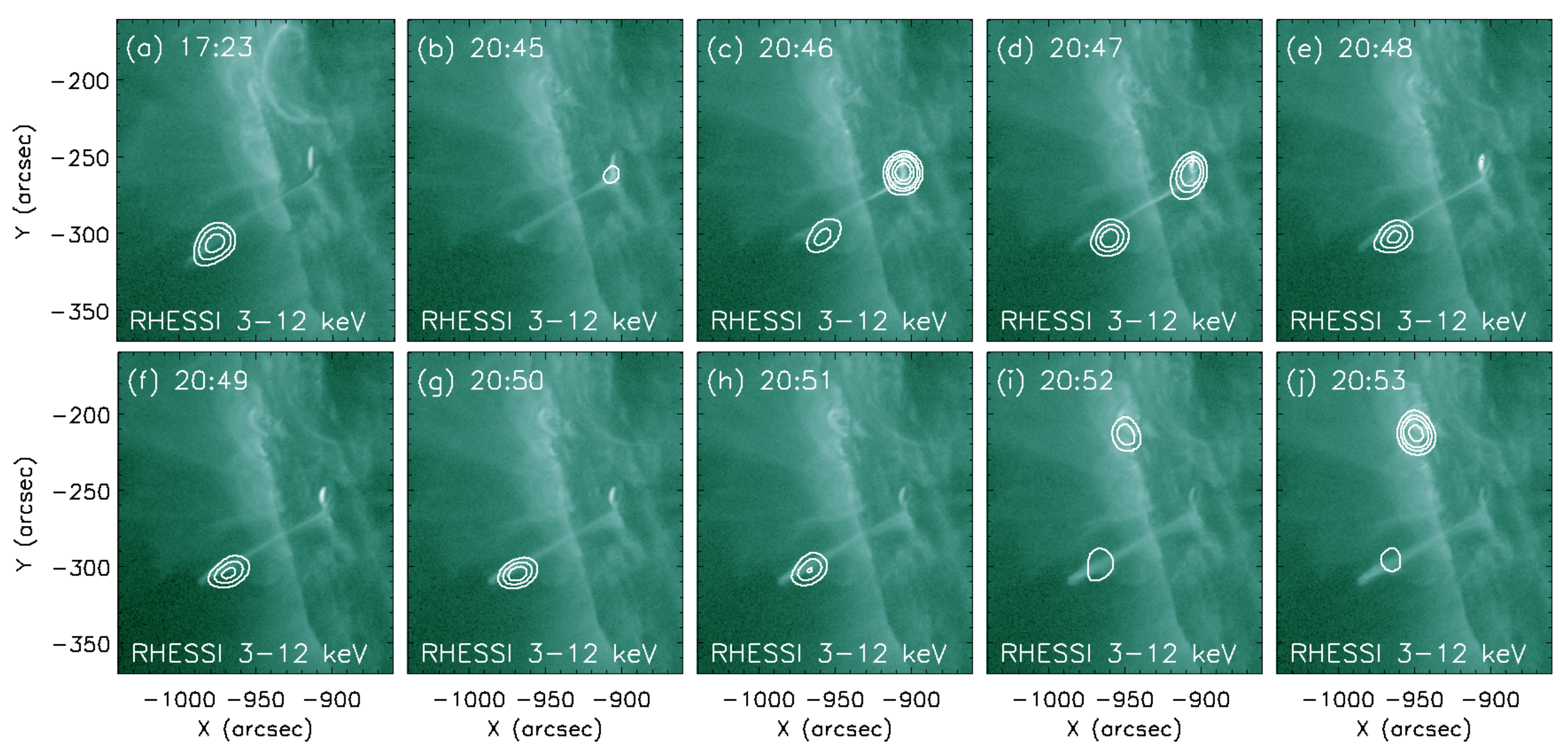}
	\caption{{\rhessi} contours in 3-12 keV overlaid on the AIA 94 {\AA} images. Panel (a) shows an image of the earlier event and panels (b)-(j) show time slices of the later event. HXR emissions were observed near the top of the jet in both events. The later event showed three different HXR sources: one at the base of the jet at $\sim$20:46 UT (panel c), one near the top of the jet at $\sim$20:50 UT (panel g), and one to the north of the jet at $\sim$20:53 UT (panel j).}
	\label{fig:rhessi_im}	
\end{figure*}

{\rhessi} was in eclipse during 16:46-17:23 UT, so it didn’t capture the entire first jet; but {\rhessi} did have full coverage for the later jet. Figure \ref{fig:rhessi_im} shows {\rhessi} images in 3-12 keV using detectors 3, 6, 8, and 9. All images were produced using the CLEAN algorithm in the HESSI IDL software package. In both events, HXR emissions were observed near the top of the jet. Furthermore, time slices of the later jet show that there were actually three HXR sources in that event. The first HXR source appeared at the base of the jet a few minutes after the jet’s starting time and peaked at around 20:46 UT. The location of this source is consistent with the erupting minifilament site where magnetic reconnection took place. Meanwhile, starting from $\sim$20:46 UT, the second HXR source appeared near the top of the jet and it became dominant during 20:48-20:51 UT. Finally, after the source at the jet top had faded away, another HXR source was observed to the north of the jet which reached its maximum intensity at $\sim$20:53 UT.

\subsection{XRT data} \label{sec:xrt}
XRT provides additional coverage of high-temperature plasma beyond AIA hot channels and {\rhessi}, though only data in the thin-Be filter for part of each jet are available. This filter is sensitive to plasma temperatures around 10 MK, and shows very similar jet behavior as the AIA 94 {\AA} and 131 {\AA} filters. Here we include these data as supplementary observations.

Thin-Be filter data were available for the first 15 minutes of the earlier jet (before 17:30 UT) with a cadence of a half minute. The jet started with a very fast flow at $\sim$17:15 UT, and reached its maximum extent in just a few minutes. Then after $\sim$17:20 UT, it grew slightly wider and remained visible towards the end of the observation time (Figure \ref{fig:xrt_iris}). As for the later jet, XRT missed most of its erupting process since no data were available between 20:45 UT and 21:09 UT. But after 21:09 UT, the jet was still visible in the thin-Be filter until it finally faded away at $\sim$22:00 UT (not shown).

\subsection{IRIS data} 
\begin{figure*}
	\centering
	\includegraphics[width=0.8\textwidth]{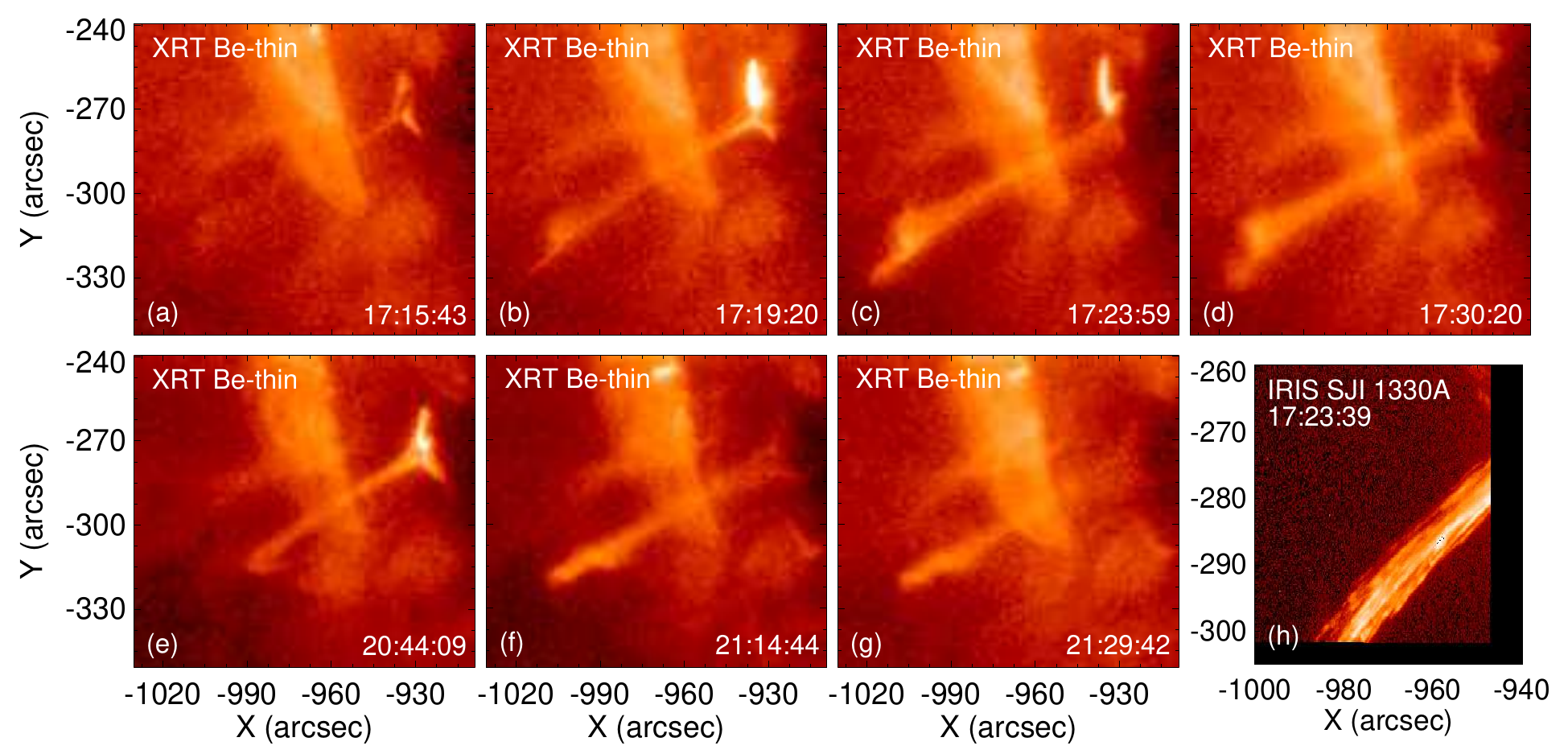}
	\caption{XRT and {\iris} images of the two jets at selected times. Panels (a)-(d): XRT Be-thin images of the earlier event. Jet evolution is similar to that in AIA hot filters (94 {\AA} and 131 {\AA}). Panels (e)-(g): XRT Be-thin images of the later event. No data was available between 20:44 and 21:09 UT. Panel (h): An {\iris} SJI 1330{\AA} image of the earlier event. The jet was located at the corner of the {\iris} field of view. }
	\label{fig:xrt_iris}	
\end{figure*}

{\iris} has full temporal coverage and partial spatial coverage of the earlier jet in its 1330 {\AA} slit-jaw images. This channel is sensitive to temperatures around 0.02 MK with a spatial resolution of 0.33 arcsec and a cadence of 10 seconds, thus it helps to investigate the dynamics of plasma at chromospheric temperatures. For the earlier event, the jet was at the corner of the field of view and most part of the jet body (but no jet base or top) was captured in these images (Figure \ref{fig:xrt_iris}). Jet evolution in this channel was similar to that in the AIA 304 {\AA} channel, and we used these data (in addition to the AIA data) to estimate jet velocities (see Section \ref{velocities}). However, this channel had much less coverage of the later jet and it was not considered for that event.

\section{Data analysis} \label{sec:analysis}

\subsection{Differential emission measure (DEM) analysis} \label{sec:DEM}
We carried out a differential emission measure (DEM) analysis to investigate the temperature profile for these two events. A DEM describes a plasma distribution with respect to temperature along the line of sight, and it is directly related to the observed flux $F$ for a particular instrument via
\begin{equation}
	F=\int{R(T) \cdot DEM(T) \, dT},
\end{equation} 
where $R$ is the temperature response of that instrument. In this analysis, we used the regularization method developed by \citet{2012A&A...539A.146H, 2013A&A...553A..10H} for DEM inversion. We considered two different data selections: (a) AIA bandpass filter data only, where we used data from six AIA bandpass filters that are sensitive to coronal temperatures: 94 {\AA}, 131 {\AA}, 171 {\AA}, 193 {\AA}, 211 {\AA}, and 335 {\AA}; and (b) a combination of multi-instrument data, where we used the same set of AIA data, together with HXR measurements in 4-5, 5-6, and 6-7 keV bands from {\rhessi}, and thin-Be filter data from XRT if available. The {\rhessi} 4-5 keV and 5-6 keV energy bands were selected because they measure plasma temperature via the bremsstrahlung continuum, and the 6-7 keV energy band was particularly important as it includes the 6.7 keV Fe line complex. The uncertainties for AIA data were estimated via the SSWIDL procedure ``\texttt{aia\textunderscore bp\textunderscore estimate\textunderscore error.pro}'', added in quadrature with a systematic error of 10\%. The uncertainties for {\rhessi} and XRT data were both estimated as 20\%.

The temperature responses for AIA and XRT filters were generated through SSWIDL routines ``\texttt{aia\textunderscore get\textunderscore response.pro}'' and ``\texttt{make\textunderscore xrt\textunderscore temp\textunderscore resp.pro}'', respectively. To obtain the temperature responses for {\rhessi} in different energy bands, we first calculated the isothermal HXR spectra as a function of energy for multiple temperatures ranging from 3 MK to 30 MK, using the SSWIDL routine ``\texttt{f\textunderscore vth.pro}''. Thus for each energy band, we obtained a series of photon fluxes at different temperatures, which would correspond to the temperature response (in photon space) for that energy band after applying proper normalization. (The {\rhessi} instrument response was already taken into account when producing HXR images, thus it did not need to be included in the temperature response.) In above calculation, coronal abundances were adopted.

\begin{figure}
	\plotone{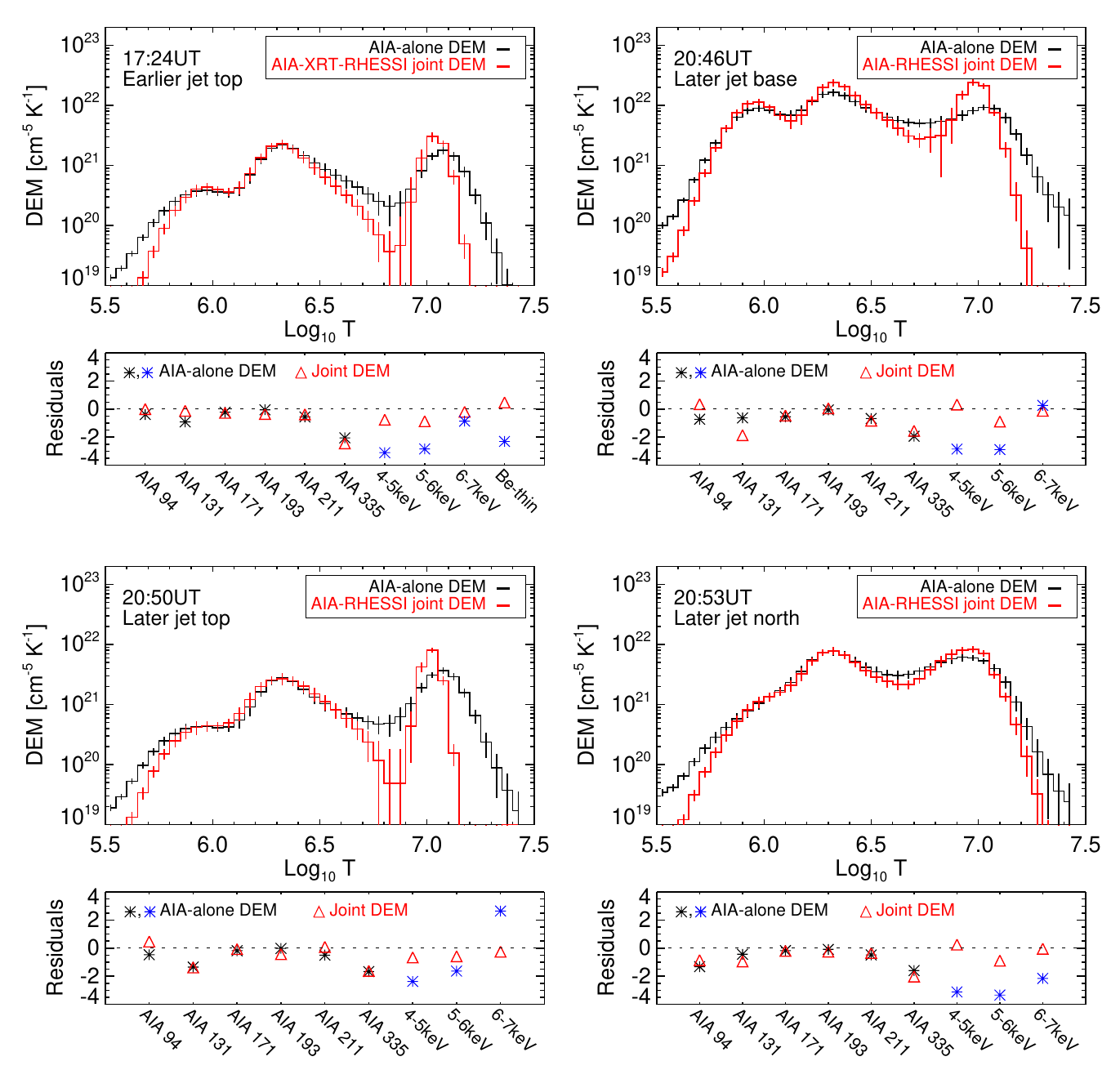}
	\caption{DEMs and data-space residuals (defined as (model-data)/error) for four different selected regions: the top of the earlier jet at 17:24 UT (top left), the base of the later jet at 20:47 UT (top right), the top of the later jet at 20:49 UT (bottom left), and a loop to the north of the later jet at 20:53 UT (bottom right). The black lines show results for the DEM inversion that used AIA data only, and the red lines show results for the DEM inversion that used multi-instrument data from AIA, {\rhessi}, and XRT if available. The AIA-alone DEMs and the joint DEMs agree qualitatively, but the joint DEMs require a more isothermal and slightly cooler high-temperature component for each source. In the residual plots, asterisks stand for results from the AIA-alone DEMs (among which black asterisks show the residuals in AIA channels that were used for the DEM inversion while blue asterisks show the residuals in {\rhessi} energy bands and possibly XRT Be-thin filter predicted by this DEM), and red triangles stand for results from the joint DEMs. The HXR fluxes predicted by the joint DEMs have a much better agreement with the actual data.}
	\label{fig:dem_ave}	
\end{figure}

We calculated the DEMs for four regions where HXR emissions were observed, around the times when each HXR source reached its maximum intensity: the top of the earlier jet at 17:24 UT, the base of the later jet at 20:46 UT, the top of the later jet at 20:50 UT, and the loop to the north of the later jet at 20:53 UT. Each region was selected based on contours in AIA 131 {\AA} images and the observed intensities were averaged over the whole region. We first obtained the DEM results using AIA data only (black lines in Figure \ref{fig:dem_ave}), as well as the corresponding residuals in data space (asterisks in Figure \ref{fig:dem_ave}). All these AIA-alone DEMs indicate the existence of multi-thermal plasma, each with a high-temperature component peaking around 10 MK. However, although these AIA-alone DEMs had good enough predictions in the AIA channels that were used for the DEM inversion, they failed to predict the HXR measurements well. As shown by the blue asterisks in the residual plots, the photon fluxes predicted by the AIA-alone DEMs in the {\rhessi} 4-5 keV and 5-6 keV energy bins were always lower than the actual measurements, and the line emissions in the 6-7 keV energy bin were very prominent compared to the bremsstrahlung continuum.

\begin{figure}
	\centering
	\includegraphics[width=0.8\textwidth]{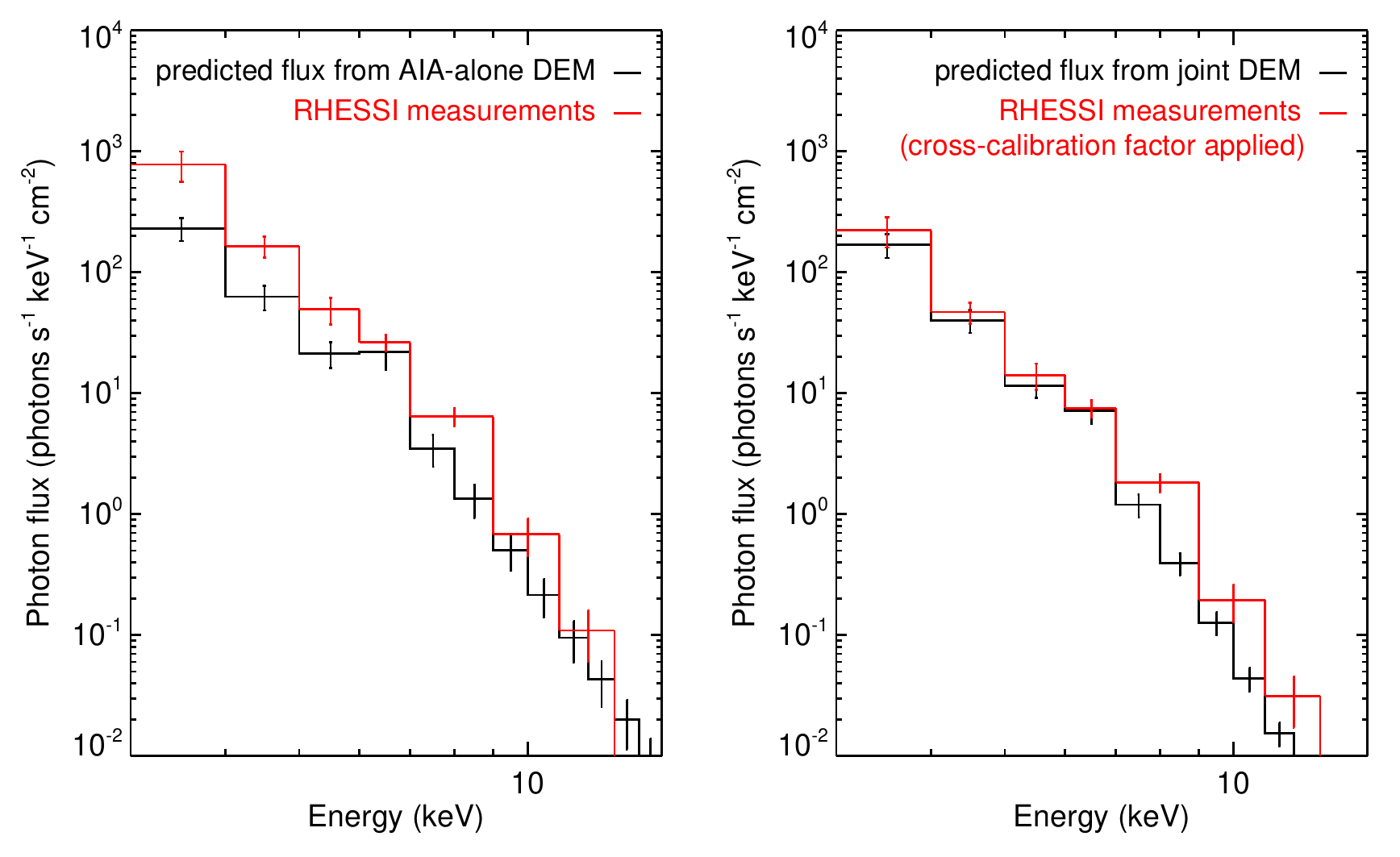}
	\caption{\textit{Left}: HXR spectrum for the source at the top of the earlier jet deduced from the AIA-alone DEM (black), compared to {\rhessi} measurements (red). \textit{Right}: HXR spectrum for the same source deduced from the joint DEM (black), compared to {\rhessi} measurements with a cross-calibration factor applied (red). With a cross-calibration factor of 3.5, the HXR spectrum from the joint DEM could successfully predict the bremsstrahlung continuum and the line feature at 6.7 keV simultaneously.}
	\label{fig:comp}	
\end{figure}

To have a better understanding of the level of agreement between AIA and {\rhessi}, we carried out a more complete quantitative comparison of HXR fluxes between these two instruments. In this exercise, we predicted the HXR spectrum in the 3-15 keV energy range for the top region of the earlier jet using the AIA-alone DEM and compared it with the spectrum directly measured by {\rhessi}. The predicted HXR spectrum was calculated according to Eq.(1), with R being the {\rhessi} temperature response. The results are shown in the left panel of Figure \ref{fig:comp}, where the AIA-predicted HXR fluxes were consistently lower than the {\rhessi} fluxes, indicating a possible cross-calibration factor between AIA and {\rhessi}. And again, the AIA-alone DEM predicted much stronger line emissions in the 6-7 keV energy bin over the continuum as compared to the actual {\rhessi} observation. Reasons for these disagreements could be some instrumental effects that are not well understood, such as the change in {\rhessi} blanketing with respect to time, or could be the possible ``non-standard'' elemental abundances in these events that we are unable to characterize.

Because of the discrepancies mentioned above, incorporating {\rhessi} data into this DEM analysis is challenging. To obtain a DEM solution that could successfully predict both the HXR continuum and the line feature at the same time, we found that a cross-calibration factor between AIA and {\rhessi} was required and the initial DEM guess must be very carefully chosen. We had the best chance of success when using a ``modified'' AIA-alone DEM as the initial guess, where we substituted the high-temperature component of each AIA-alone DEM with a Gaussian distribution that peaked around the temperature given by {\rhessi} spectroscopy (more details of spectroscopy will be discussed in Section \ref{sec:imag_spec}). The height, width, and exact peak location of that Gaussian distribution were tested with a series of values, and we selected the most robust ones. In this piece of analysis, we scaled down the photon fluxes from {\rhessi} by a factor of 3.5, but this cross-calibration factor could be in the range of 3-5 and was not well constrained. Incidentally, this factor that we found here is similar to the AIA-{\rhessi} discrepancy found by \citet{2013ApJ...779..107B}. In addition, in some literature a factor of 2-3 was suggested for cross-calibration between AIA and XRT \citep[e.g.][]{2015ApJ...806..232S, 2017ApJ...844..132W}. We found that multiplying a factor of 2 to the XRT response would result in a better agreement between the predicted and measured Be-thin filter data; thus that factor was also included here.

The joint DEMs (red lines) and the data-space residuals (red triangles) are also plotted in Figure \ref{fig:dem_ave}, along with the AIA-alone DEM results. These joint DEMs are the only set of solutions we found that fit both the line emission and the bremsstrahlung continuum well. For all the selected regions, the joint DEMs have very similar cool components as the AIA-alone DEMs, but the hot components of the joint DEMs tend to be more isothermal and slightly cooler. Particularly, HXR constraints significantly reduced the amount of plasma above $\sim$15 MK (otherwise the predicted line emission was always too prominent). However, previous studies have seen larger discrepancies between bandpass filter DEMs and the ones that included HXR constraints. For example, in the DEM analysis for a quiescent active region presented by \citet{2009ApJ...704..863S}, a high-temperature component that peaked around $10^{7.4}$K was found when only using data from XRT filters, but the DEM for that component was reduced by more than one order of magnitude when combining observations from both XRT and {\rhessi}. Compared to that study, the AIA-alone DEMs here are not too far from the joint DEMs that incorporated HXR data.

We further compared the HXR fluxes predicted by the joint DEM with {\rhessi} measurements (with a cross-calibration factor applied) for the top region of the earlier jet, as shown in the right panel of Figure \ref{fig:comp}. As expected, the two HXR spectra had a much better agreement at lower energies than the spectra predicted from AIA-only DEMs did, including both the overall continuum and the line feature. Besides, for higher energies around 10 keV, {\rhessi} measurements had systematically higher emissions, suggesting a possible non-thermal component for this source. This is consistent with our later findings through spectral analysis (Section \ref{sec:imag_spec}). As a side note, later spectral analysis suggests that for some of the HXR sources non-thermal emissions might dominate in the 6-7 keV (and maybe 5-6 keV) energy bin(s). In this scenario, the fluxes from those thermal sources would be even lower and the joint-DEMs shown here provide upper limits for the possible amount of hot plasma. 

\begin{figure}
	\plotone{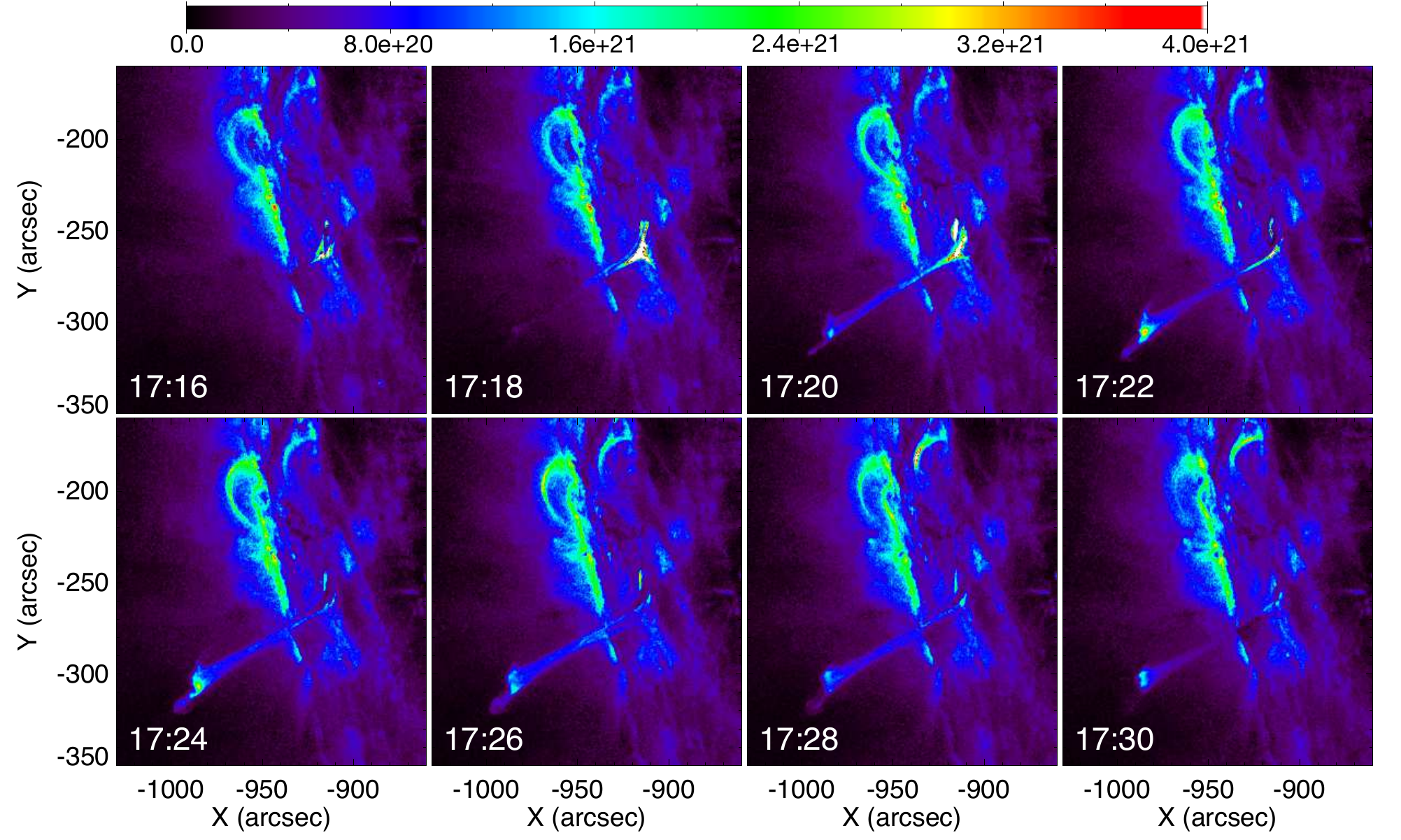}
	\caption{Temporal evolution of the DEM maps in 11-14 MK for the earlier jet. Hot plasma appeared at the base of the jet during the first few minutes, but starting from $\sim$17:20 more and more hot plasma was observed near the top of the jet and the top source became dominant at $\sim$17:22. Color scale is in units of cm$^{-5}$ K$^{-1}$.}
	\label{fig:em1}	
\end{figure}

\begin{figure}
	\plotone{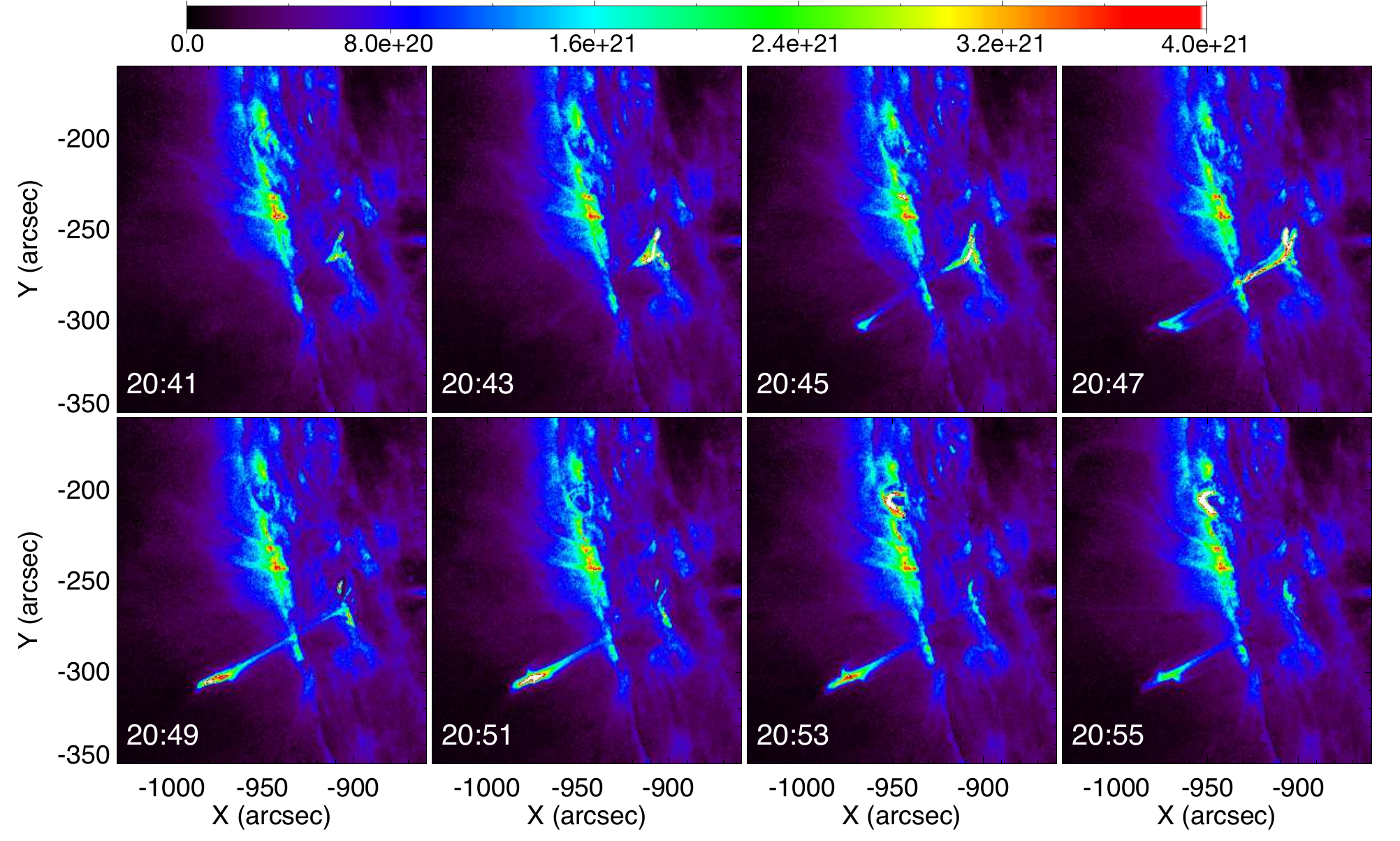}
	\caption{Temporal evolution of the DEM maps in 11-14 MK for the later jet. Similar to the earlier event, hot plasma first appeared at the base of the jet, but was also observed near the top of the jet starting from $\sim$20:45 and the top source became dominant at $\sim$20:50. Color scale is in units of cm$^{-5}$ K$^{-1}$.}
	\label{fig:em2}	
\end{figure}

As the last part of the DEM analysis, we examined the temporal evolution of the DEM maps in 11-14 MK (i.e. the hot component) for each event (Figures \ref{fig:em1} and \ref{fig:em2}). Because of the missing {\rhessi} and XRT data, and the fact that the AIA-alone DEMs in this temperature range qualitatively agree with the joint DEMs, these DEM maps were all generated using AIA data only. In both events, hot plasma first appeared at the base of the jet (which was the same location from where the minifilament erupted and magnetic reconnection occurred); however, as the hot plasma at the jet base gradually cooled down, more and more hot plasma was observed near the top of the jet and that location was mostly stationary. These DEM maps show consistent results with the location and temporal evolution of the {\rhessi} HXR sources.

\subsection{Jet velocities} \label{velocities}

Identifying different velocities associated with the jet will be helpful to differentiate possible mechanisms behind those jets. A common method for velocity estimation is making time-distance plots \citep[e.g.][]{2016A&A...589A..79M, 2020ApJ...889..183M}. Such plots are usually produced by putting together time slices of the intensity profile along the direction of the jet, in which case the jet velocities (in the plane of the sky) are the slopes. To take into account everything within the width of the jets, here we selected a rectangular region around each jet and we summed the intensities across the width of this region. 

\begin{figure}
	\centering
	\includegraphics[width=0.75\textwidth]{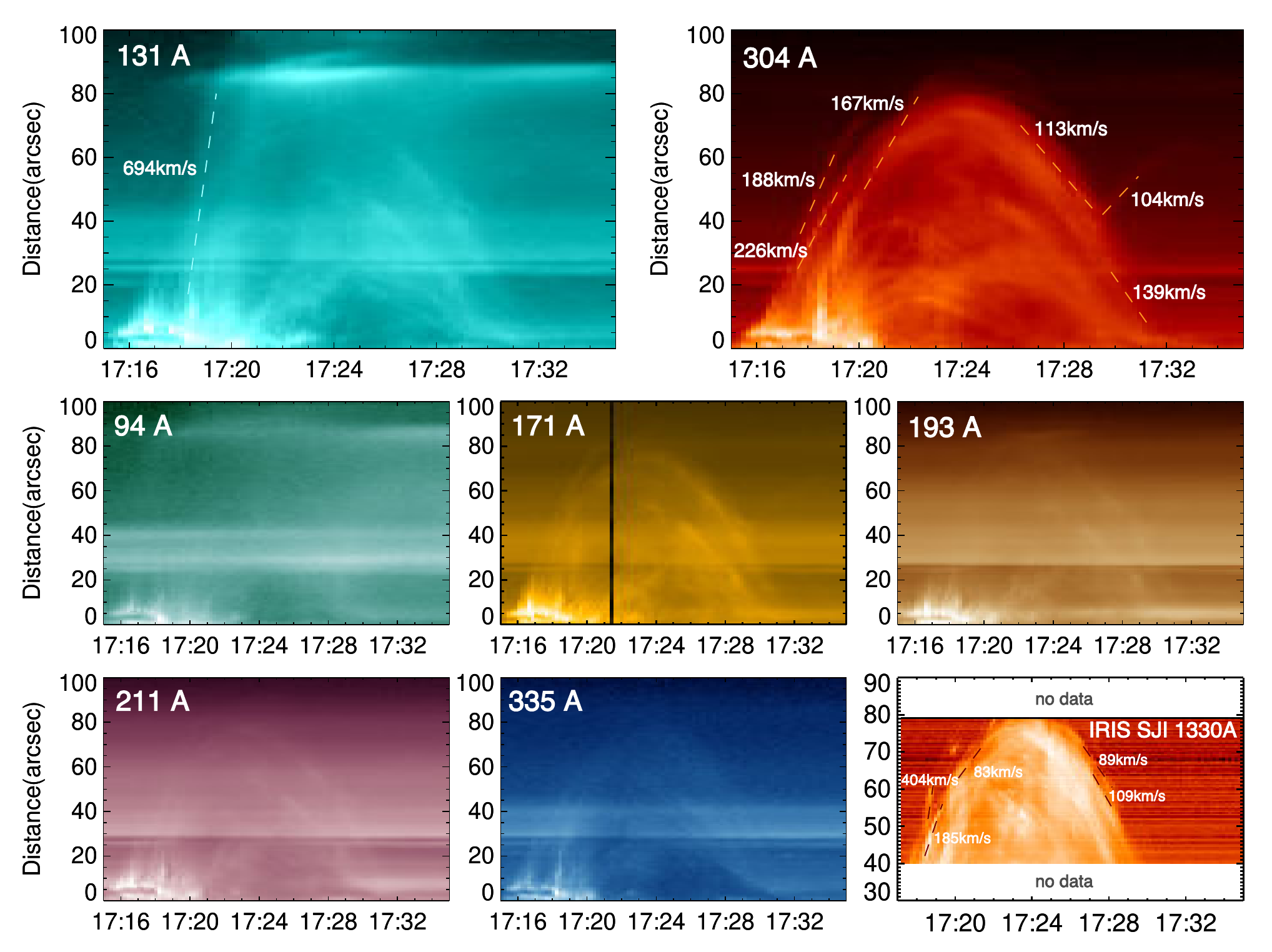}
	\caption{Time-distance plots of the earlier jet in seven AIA EUV filters and the {\iris} 1330 {\AA} filter. Slopes for velocity calculation are shown as dashed lines. The chromospheric filters (AIA 304 {\AA} and {\iris} 1330 {\AA}) show various upward velocities around 200 km/s and downward velocities around 110 km/s. The AIA 131 {\AA} filter (sensitive to both $\sim$0.4 MK and $\sim$10 MK) shows a faster upward velocity of 694 km/s at the beginning of the jet. In the rest of the AIA filters, velocities are less apparent. (The black line at $\sim$17:21 in the  171 {\AA} plot is due to some instrument issue.)}
	\label{fig:td1}	
\end{figure}

Figure \ref{fig:td1} shows the time-distance plots for the earlier jet using seven EUV filters of AIA and the slit-jaw 1330{\AA} filter of {\iris}. Interestingly, the chromospheric filters (AIA 304 {\AA} and {\iris} 1330 {\AA}) are the ones where the velocities are most clearly identified and show very consistent results. The 304 {\AA} filter shows multiple upward velocities ranging from 104 km/s to 226 km/s, while the 1330 {\AA} filter shows upward velocities ranging from 83 km/s to 404 km/s (the uncertainties for those velocities are on the order of 10-20$\%$ considering the pixel size and the temporal cadence of the images). Also, both filters clearly indicate that some plasma returned to the solar surface (possibly) along the same trajectory as the original jet, with downflow velocities of $\sim$110 km/s. For the rest of the AIA filters, similar upward and downward velocities as mentioned above can partly be seen in the 171 {\AA} filter (sensitive to $\sim$0.6 MK plasma), but could barely be seen in other ones. However, there were also some really fast outflows at the beginning of this jet in the 131{\AA} filter (sensitive to both $\sim$0.4 MK and $\sim$10 MK plasma), which has a velocity of $\sim$700 km/s.

\begin{figure}
	\centering
	\includegraphics[width=0.75\textwidth]{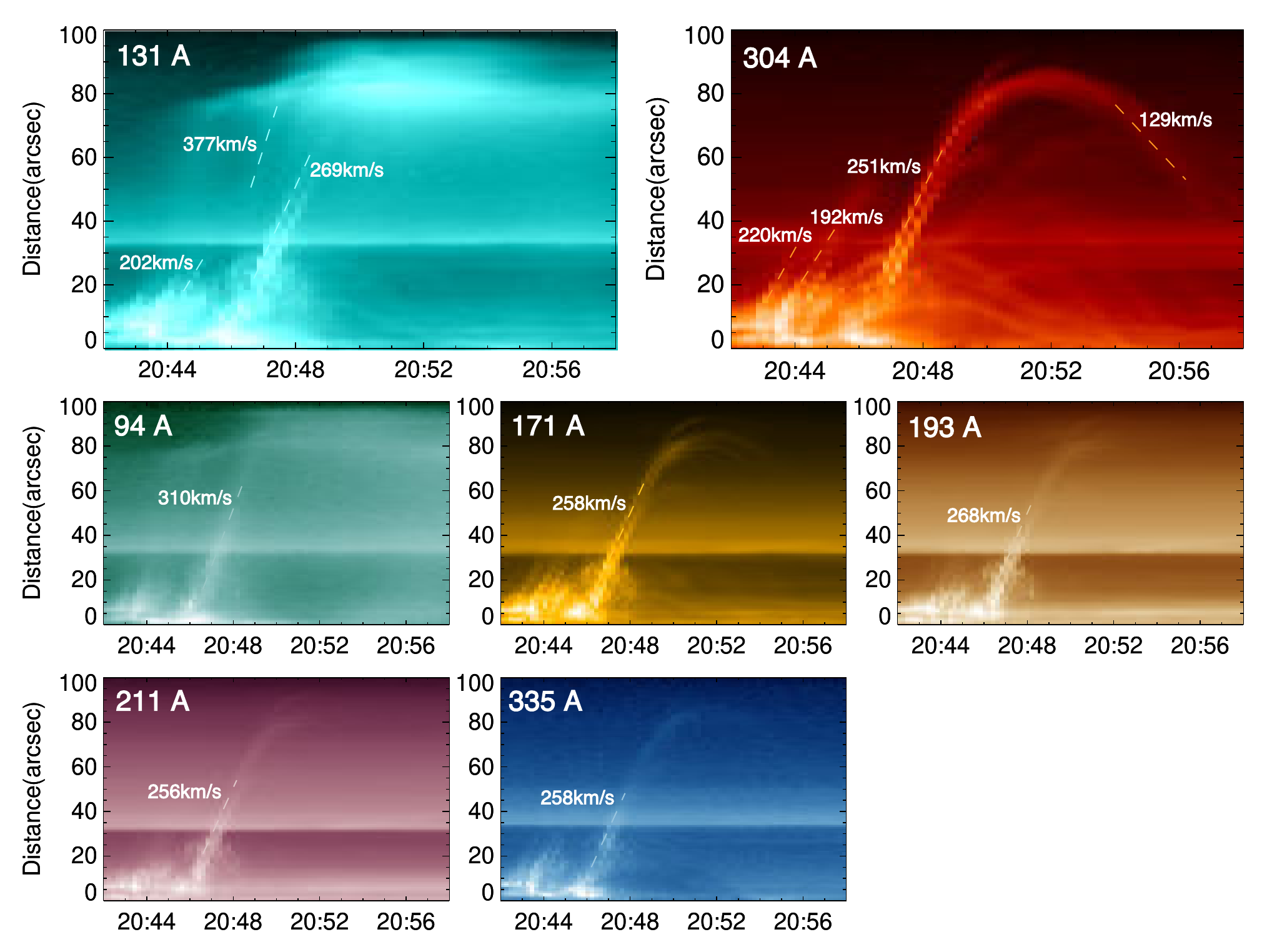}
	\caption{Time-distance plots of the later jet in seven AIA EUV filters. Slopes for velocity calculation are shown as dashed lines. Unlike the earlier event, the main upward velocities can be clearly identified in all filters. The AIA 131 {\AA} filter still shows a faster upward velocity of 377 km/s at the beginning of the jet, but this velocity is less apparent.}
	\label{fig:td2}	
\end{figure}

The time-distance plots for the later jet describe a slightly different picture (Figure \ref{fig:td2}). The 304 {\AA} filter again shows multiple upward velocities ranging from 192 km/s to 251 km/s, and downward velocities around 130 km/s (the uncertainty is on the order of 10-20\%). However, these main upward velocities can be clearly identified in all seven AIA filters, including both cool and hot ones. Besides, in the 131 {\AA} filter, a faster outflow at the beginning of the jet is still identifiable but harder to see compared to the earlier event, and the velocity for this outflow is 377 km/s. These velocities will be compared to other studies and to models in Section \ref{sec:v_discuss}.

\subsection{Imaging spectroscopy} \label{sec:imag_spec}

To study the accelerated electron populations in these events, we performed imaging spectroscopy for the four HXR sources observed by {\rhessi}. For each source, a one-minute time interval during which the source reached its maximum HXR intensity was first selected by eye based on the {\rhessi} images. These images were produced using the CLEAN algorithm and detectors 3, 6, 8, and 9. Then we chose a circular region which contained that source and obtained the spectrum for the selected region. Finally, we carried out spectral fitting using the OSPEX software package in the energy range of 3-15 keV.  

\begin{figure}
	\plotone{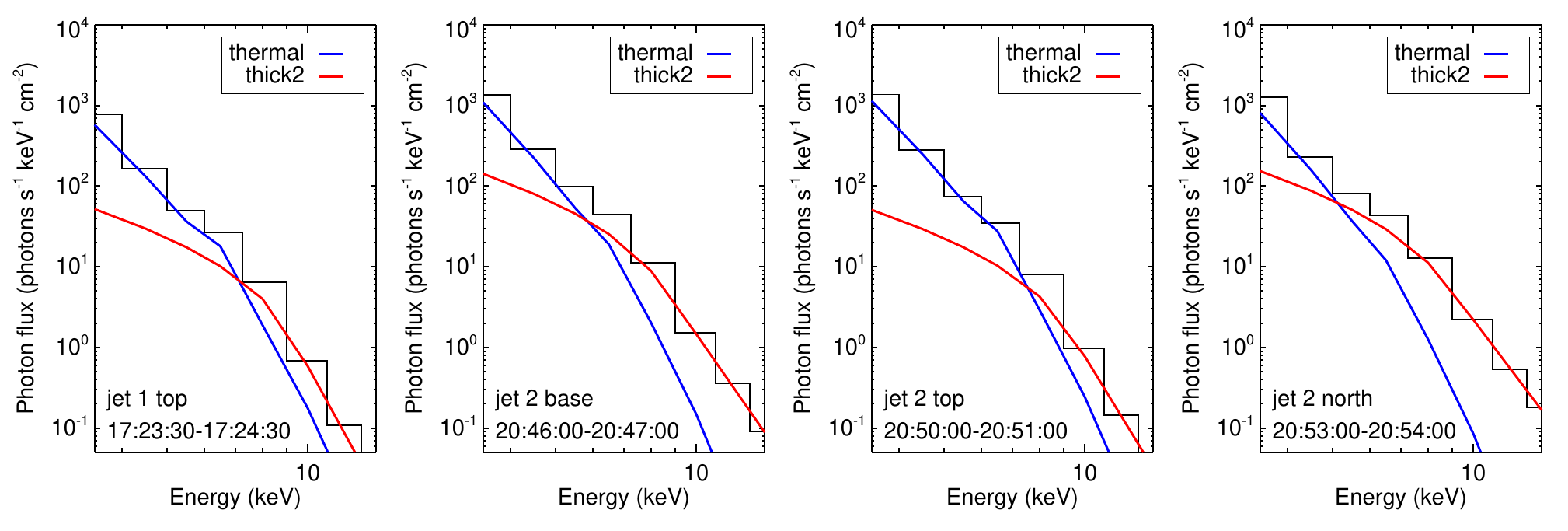}
	\caption{{\rhessi} spectra for the four HXR sources observed in these two events, each during a one-minute interval when the source approximately reached its maximum HXR intensity. All spectra can be fitted well with an isothermal (blue) plus thick-target (red) model. Note that in these fits, the isothermal temperatures were fixed to be the average temperatures of the hot components derived from the joint DEMs.}
	\label{fig:spec}	
\end{figure}

As mentioned in Section \ref{sec:DEM}, the comparison of the joint DEM and {\rhessi} measurements suggests a non-thermal component in the HXR spectrum. To further confirm this, we first fitted the spectra with an isothermal model (not shown), but the models always overpredicted the fluxes at the 6.7 keV line complex and had systematically low emissions at energies above 10 keV for all the sources. This indicated that there should be another component in the spectra, either due to a second thermal distribution or a non-thermal distribution. However, the results of fitting for a double thermal model (not shown) had unphysical fit parameters for one of the thermal components, and it again overpredicted the line emission, making this scenario unlikely. Therefore, we confirmed that there should be non-thermal emissions in these events. We then added a thick-target non-thermal component to the fitting (the justification for the thick-target regime will be discussed in Section 4.2), and could obtain good fits across the entire observed energy range. (We used the temperatures from those fits as a reference when generating the initial DEM guess for the joint DEM inversion.) However, due to the limited number of energy bins and the number of free parameters, some of the fit parameters were not well constrained and had uncertainties over 100$\%$. To further reduce the uncertainties, we performed another fitting with a fixed temperature, which was chosen to be the average temperature of the hot component derived from the joint DEMs. The resulting spectra are shown in Figure \ref{fig:spec} and the parameters are reported in Table \ref{tab:spec_fit}. Interestingly, the non-thermal electron power laws in all sources have similar spectral indices around 10 and low energy cutoffs around 9 keV. While these non-thermal power laws are steeper than in most flares, the parameters are consistent with the range found for microflares by \citet{2008ApJ...677..704H}.

\begin{deluxetable*}{lcccccc}
	\tablenum{1}
	\tablecaption{Fit parameters for the four {\rhessi} HXR sources assuming an isothermal plus thick-target model \label{tab:spec_fit}}
	\tablehead{
		\colhead{} & \colhead{Time} & 
		\colhead{Emission measure} &
		\colhead{Temperature} & \colhead{spectral index} &
		\colhead{low energy cutoff} \\
		\colhead{} & \colhead{} &
		\colhead{($10^{46} \mathrm{cm}^{-3}$)} &
		\colhead{(MK, fixed)} & \colhead{} & \colhead{(keV)} 
	}
	\startdata
	Jet 1 top source & 17:23:30-17:24:30 & 1.8 $\pm$ 0.5  & 11.1 & 11.4 $\pm$ 3.6 & 9.3 $\pm$ 1.6 \\
	Jet 2 base source & 20:46:00-20:47:00 & 4.7 $\pm$ 1.0 & 9.9 & 9.1 $\pm$ 0.9 & 8.5 $\pm$ 0.9 \\
	Jet 2 top source & 20:50:00-20:51:00 & 4.1 $\pm$ 0.5 & 10.5 & 10.3 $\pm$ 1.4 & 9.4 $\pm$ 1.2 \\
	Jet 2 northern source & 20:53:00-20:54:00 & 3.9 $\pm$ 1.3 & 9.6 & 8.5 $\pm$ 0.8 & 8.7 $\pm$ 1.0 
	\enddata
\end{deluxetable*}

\section{Discussion} \label{sec:discussion}
\subsection{Jet velocities and driving mechanisms} \label{sec:v_discuss}
In these two events, we observed two types of upward velocities in jets. One type falls within the range of 80-400 km/s, most clearly seen in the AIA 304 {\AA} and {\iris} 1330 {\AA} filters (sensitive to chromospheric temperatures) and possibly visible in other filters. This type of velocity is consistent with many previous studies of coronal jets \citep[e.g.][]{1996PASJ...48..123S, 2007PASJ...59S.771S, 2016A&A...589A..79M, 2016ApJ...832L...7P, 2020ApJ...889..183M} where the jet velocities usually range from a few tens of km/s to $\sim$500 km/s, with an average around 200 km/s. The other velocities, $\sim$700 km/s and $\sim$400 km/s respectively for the two jets, could only be identified in the 131 {\AA} filter (sensitive to $\sim$0.4 MK and hot temperatures $\sim$10 MK) at the beginning of each event (though harder for the later jet). The velocity of $\sim$400 km/s is still within the common range for coronal jets, but it is faster than the other velocities observed in that jet. The velocity of $\sim$700 km/s seems to be faster than most of the observed jets. However, such velocities are not rare and have been reported in a few coronal jet observations by XRT \citep{2007PASJ...59S.771S, 2007Sci...318.1580C}.

One possible acceleration mechanism for coronal jets is chromospheric evaporation, which is also the responsible mechanism for some plasma flows in solar flares. In this process, the energy released from magnetic reconnection is deposited in the chromosphere, compresses and heats the plasma there, and produces a pressure-driven evaporation outflow on the order of sound speed. In fact, \citet{1984ApJ...281L..79F} derived a theoretical upper limit for the velocity of this evaporation outflow to be 2.35 times the local sound speed, where the sound speed $c_s$ can be calculated as: $c_{s}=147\sqrt{\frac{T}{\mathrm{1MK}}}$ km/s assuming an isothermal model \citep[e.g.][]{2004psci.book.....A}. In the 304 {\AA} filter (characteristic temperature $10^{4.7}$ K), this upper limit corresponds to a very low speed of 77 km/s, indicating that the cool plasma is unlikely driven by chromospheric evaporation. Also, common velocities reported by observations of chromospheric evaporation usually fall within the range of tens of km/s up to 400 km/s \citep[e.g.][]{2013ApJ...767...55D, 2015ApJ...811..139T, 2015ApJ...805..167S}, thus chromospheric evaporation seems not able to explain the very fast flow of the earlier jet observed in the hot 131{\AA} filter. Furthermore, it is expected that the velocity would increase with the temperature if a jet is generated by chromospheric evaporation \citep{2012ApJ...759...15M}, but here we have seen very consistent velocities in all seven AIA filters that are sensitive to different temperatures in the later event. For all these reasons, if both jets are driven by the same mechanism, that mechanism is likely \textit{not} chromospheric evaporation but magnetic tension instead. However, it is not clear why the earlier jet shows more complicated and various velocities (even in a single channel) if both jets are driven similarly.

\subsection{Particle acceleration locations}  
In Section \ref{sec:imag_spec}, we fitted the {\rhessi} spectra of the four HXR sources with an isothermal plus thick-target model. Here we first justify that the thick-target regime is a reasonable approximation.

\begin{figure}
	\centering
	\includegraphics[width=0.75\textwidth]{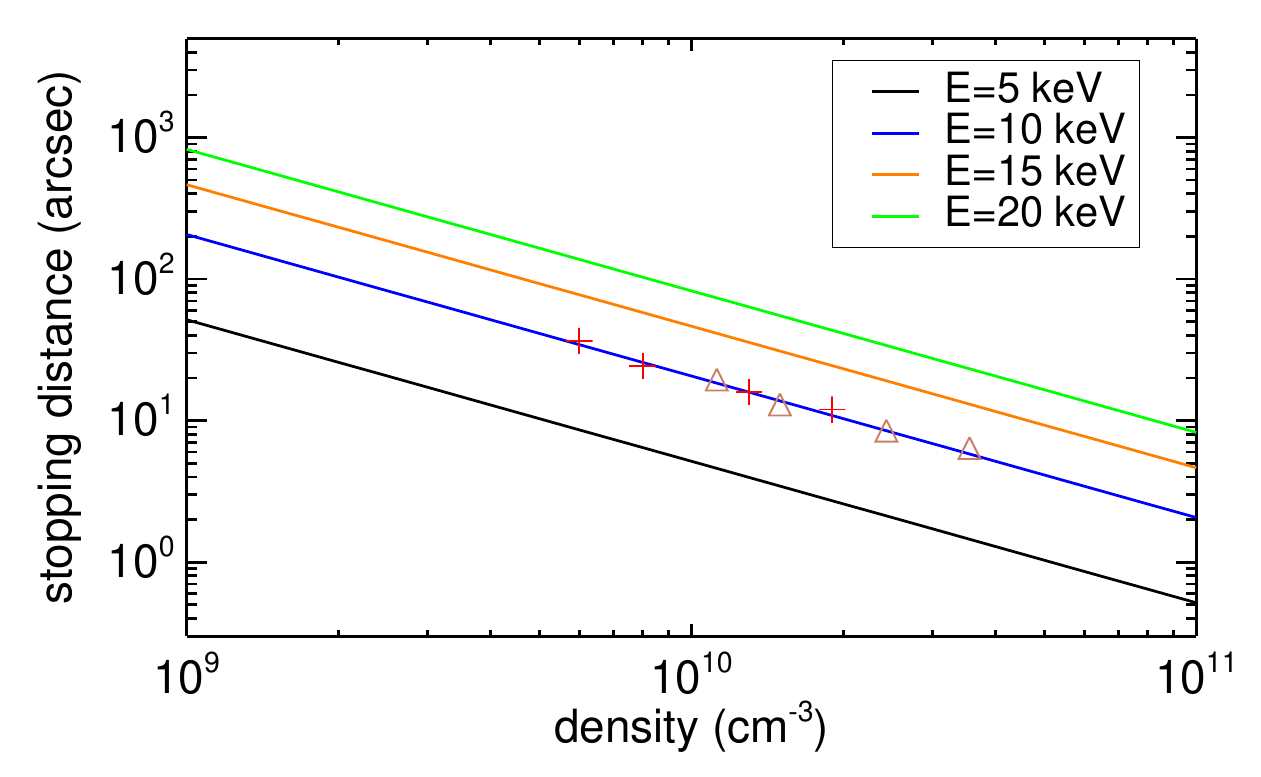}
	\caption{The relation between collisional stopping distances and ambient plasma densities for electrons of certain energies. Red plus signs mark the values for the densities (without a cross-calibration factor) and average energies for the four observed HXR sources, while brown triangles mark the values with a cross-calibration factor applied. The stopping distances for these sources are less than a few tens of arcsecs, but for lower densities and/or higher electron energies, accelerated electrons could travel an appreciable distance in the corona. }
	\label{fig:stopping_d}	
\end{figure}

The column depth (defined as $N_s=\int ndz$ where n is the plasma density) to fully stop an electron of energy E (in units of keV) can be calculated as: $N_s=1.5\times10^{17}\mathrm{cm^{-2}} E^2$ \citep[e.g.][]{krucker2008hard}. Based on this formula, Figure \ref{fig:stopping_d} plots the relation between the stopping distance and the plasma density for a given electron energy. Under the thick-target regime, according to Table \ref{tab:spec_fit}, the average electron energy for the source at the top of the earlier jet is $\sim$10 keV and the density there is $6\times10^{9}$ $\mathrm{cm}^{-3}$ (derived from the joint DEM), which corresponds to a distance of 36 arcsec in average that the electrons can travel before fully stopped by the ambient plasma. Similar average electron energies around 10 keV are found for the other three HXR sources in the later event, and the densities of those sources are $(0.8-1.9)\times10^{10}$ $\mathrm{cm}^{-3}$, resulting in stopping distances of 10-28 arcsec. Moreover, in Section \ref{sec:DEM}, we report a possible cross-calibration factor around 3.5 between AIA and {\rhessi}. That factor is not included in the above calculation; however, if the cross-calibration factor is included, all the densities above would be multiplied by $\sqrt{3.5}$, corresponding to even shorter stopping distances of 5-20 arcsec. In general, these stopping distances are comparable to the size of the HXR sources, meaning that accelerated electrons deposit a considerable portion of their energies into each source. Furthermore, if this is a thin-target regime, the spectral indices would be slightly smaller but the average electron energies would still be $\sim$10 keV. This would result in very similar stopping distances that are comparable to source sizes, which is not consistent with the thin-target assumption. Therefore, we conclude that the HXR sources observed in these events can be approximated as thick targets, and mildly accelerated electrons are found at all these locations.

Since the three HXR sources observed in the later event share similar electron distributions, here comes the next question: were the HXR emissions in the later event produced by the same population of accelerated electrons that traveled to different locations, or were they produced by different groups of accelerated electrons individually? In the standard jet models, reconnection happens near the base of the jet, which would require electrons to be accelerated near the base and travel upwards along the magnetic field lines to produce the HXR source at the jet top. However, from the DEM analysis, we find that the densities in the body of the jet (where there were not many HXR emissions) are $\sim1\times10^{10}$ $\mathrm{cm}^{-3}$, so the stopping distance along the jet body is still 20-30 arcsec (for both jets). This is a few times smaller than the distance from the jet base to the jet top; therefore the HXR source at the top of each jet was produced by electrons that were accelerated very close to this source, rather than electrons that traveled far from the primary reconnection site at the jet base. This finding is in line with a similar one made for the powerful X8.3 class flare on September 10, 2017, obtained with an entirely different methodology that employs microwave imaging spectroscopy \citep{fleishman2022solar}. 

Another possible explanation for the sources at the jet top could be that the jets were actually ejected along large closed loops perpendicular to the plane of the sky rather than the so-called ``open'' field lines. Then the top of the jet is in fact the apex of the loop, which would have higher emissions purely because of the line-of-sight effect. However, even in this scenario the stopping distance along the jet body would remain the same, thus the conclusion of an additional particle acceleration site near the jet top (or loop apex) still holds regardless of jet geometry.

The HXR source to the north of the later jet appeared last among the three HXR sources, but still during a time when the jet was visible in EUV filters. It is also likely related to the jet because the formation of the jet would change the magnetic configuration of the active region, but neither the electron path nor the density along the path is clear if the energetic electrons traveled from the jet base to the northern location. The typical coronal density for an active region is about $10^{9}$ $\mathrm{cm^{-3}}$ \citep[e.g.][]{1961ApJ...133..983N}, corresponding to a stopping distance of 200 arcsec for electrons of 10 keV. Thus, in general situations energetic electrons could travel a decent distance in the corona, but it is also possible that the density in this active region is larger than that typical value. However, due to lack of data, we couldn't determine which was the case for this source.

\subsection{Energy budget}
Investigating the partition of different energy components can help to understand the energy release process in these events. Such calculations have been done in the past for a number of flares and CMEs \citep[e.g.][]{2012ApJ...759...71E, 2015ApJ...802...53A, 2016A&A...588A.116W}, but only for a few jets so far \citep{2013ApJ...776...16P}. Here we present our estimates of various energy components for the later event, including kinetic energy, gravitational energy, thermal energy, and the energy in non-thermal electrons. We calculated the maximum amount of energy that could be converted into each of the forms above.

The jet's major eruption started from 20:46 UT, which was visible in all AIA filters and had a speed of $\sim$260 km/s (Figure \ref{fig:td2}). The density of this plasma was derived from its DEM, which is $1.3\times10^{10}$ $\mathrm{cm}^{-3}$. Assuming the jet body that contained this group of plasma to be a cylinder, the peak kinetic energy of the jet is $5\times10^{26}$ erg.

The maximum height of this jet is $\sim$80 arcsec; however, as the height increases the amount of plasma that traveled there decreases, and it's not clear what fraction of plasma finally reached the maximum height. Therefore, instead of calculating the maximum gravitational energy of the jet, we set an upper limit of $2\times10^{26}$ erg, which is the gravitational energy if all the plasma of the major eruption reached a height of 80 arcsec. This upper limit is smaller than the kinetic energy of the jet, meaning the eruption is not ballistic. 

The thermal energy, $E_{th} = 3k_BTnV$, is dominated by contributions from HXR sources. Using the joint DEMs, the peak thermal energy for each HXR source is about $5\times10^{27}$ erg. This value is consistent with the flare thermal energies found for other jets in \citet{2020ApJ...889..183M}.

The energy in non-thermal electrons can be simply estimated as $E_{nonth} = N_eE_{e,ave}$ where $N_e$ is the total number of accelerated electrons and $E_{e,ave}$ is the average electron energy. Adopting the thick-target approximation and using the parameters from Table \ref{tab:spec_fit}, the non-thermal energy for each HXR source is about $(6-11)\times10^{29}$ erg. However, this value here is calculated based on {\rhessi} measurements. If we apply the cross-calibration factor between AIA and {\rhessi} to match the calculations of other energy forms, the non-thermal energy for each HXR source becomes $(3-6)\times10^{29}$ erg.

The energies of HXR sources in this event can be compared to those in previous studies of flare energetics. \citet{2012ApJ...759...71E} studied 38 eruptive events (all except one were M or X class flares and most flares were accompanied by a CME), and they found that the flare thermal energies were always smaller than the energies in accelerated particles. Similar results were found in a later study by \citet{2016A&A...588A.116W}, where the median ratio of thermal energies to non-thermal energies in electrons for 24 (C-to-X class) flares was 0.3. In a subclass of ``cold'' flares \citep{2018ApJ...856..111L},  the thermal energy is equal (within the uncertainties) to the non-thermal energy deposition \citep{2016ApJ...822...71F, 2020ApJ...890...75M, 2021ApJ...913...97F}. (Theoretically, the thermal energy cannot be less than the non-thermal energy as the latter one decays into the thermal one.) For our jet event that contains low-C class flares, the thermal energies are more than one order of magnitude smaller than the non-thermal energies. Therefore, the conclusion that the non-thermal energy is always larger than or at least equal to the thermal energy is likely consistent across a wide range of flare classes and regardless of whether the flare is associated with a jet/CME or not. 
 
However, the energy partition between the jet and the associated flares is different from the energy partition between a CME and a flare. For this event, the kinetic/gravitational energy of the jet is more than one order of magnitude smaller than the energy of the flares (thermal/non-thermal), while in \citet{2012ApJ...759...71E} the total energy of the CME is usually significantly larger. (The kinetic energy in confined flares is much smaller \citep{2021ApJ...913...97F}, though.) This variety could be explained in the minifilament eruption scenario that jets and CMEs are still both parts of the same eruptive events but the energy partition changes with scale, or this could also indicate that there are fundamental differences between jets and CMEs. To further answer this question, future studies with more samples of flare-related jets are needed.

Lastly, it should also be noted that there are still other forms of energy that were not considered in the calculations above, such as magnetic energy, wave energy, etc. These energies could also be important components of the event energy budget, but are hard to evaluate here due to limited data.

\subsection{Comparison to the current jet models} \label{comp_models}
Considering the locations of hot plasma as well as the HXR sources, these two jets are interesting examples to be compared with current jet models. On the one hand, the source at the base of the jet is consistent with what is expected from jet models. During the minifilament eruption at the jet base, magnetic reconnection happens close to the bottom of the corona, heating the plasma there directly and generating accelerated electrons near the reconnection site. The downward-traveling energetic electrons radiate bremsstrahlung emissions as they collide with the dense chromosphere, producing a HXR source and/or further heating the ambient plasma at the base of the jet. On the other hand, processes after a jet's eruption are generally not considered by those models; thus the hot plasma and the HXR source at the top of the jet are not expected. Our observations have shown that additional particle acceleration could happen at other locations besides the jet base. In other words, there could be multiple reconnection and energy release sites in a single jet event. Also, despite the significantly different particle acceleration sites (and even two separate events), the non-thermal electrons share very similar energy distributions. The spectral indices around 10 and the low energy cutoffs around 9 keV suggest that jet reconnection typically produces only mild particle acceleration. These low energy cutoffs are similar to those of the cold flares \citep[e.g.,][]{2020ApJ...890...75M}, while the spectra are much softer in the case of the jets.

Another interesting point about these events is the relation between hot and cool material. For both jets, the cool ejections observed in the 304 {\AA} filter were adjacent to the hot ejections observed in the 94 {\AA} and 131 {\AA} filters. While past simulations have successfully produced a hot jet and a cool jet (or surge) in a single event, it is generally expected that hot and cool jets are driven through different mechanisms. For example, in the simulation by \citet{1996PASJ...48..353Y}, the hot jet was accelerated by the pressure gradient while the cool surge was accelerated by magnetic tension. Similarly in an observational study by \citet{2012ApJ...759...15M}, the hot component (at coronal temperatures) was generated by chromospheric evaporation while the cool component (at chromospheric temperatures) was accelerated by magnetic force. However, though the observation of the earlier jet doesn’t conflict with this picture, the later jet had consistent velocities in hot and cool filters, indicating that some of the hot components might be driven by a very similar process as the cool components in that event. Therefore, at least in some cases the hot and cool components must be more closely related, and a jet model should be able to explain this kind of observation as well as those similar to \citet{2012ApJ...759...15M}.

\section{Summary} \label{sec:summ}
In this paper, we present a multi-wavelength analysis of two active region jets that were associated with low C-class flares on November 13, 2014. Key aspects of this study include:

\begin{enumerate}
	\item In both events, hot ($\gtrsim$10MK) plasma not only appeared near the base of the jet (which is the location of the primary reconnection site) at the beginning, but also appeared near the top of the jet after a few minutes. 
	\item Four {\rhessi} HXR sources were observed: one (at the jet top) in the first event and three (at the jet base, jet top, and a location to the north of the jet) in the later event. All those sources showed evidence of mildly accelerated electrons which had spectral indices around 10 and extended to low energies around 9 keV. 
	\item Various jet velocities were identified through time-distance plots, including major upward velocities of $\sim$250 km/s and downward velocities of $\sim$100 km/s. Fast outflows of $\sim$700 km/s or $\sim$400 km/s were observed only in the hot AIA 131 {\AA} filter at the beginning of each jet. These velocities indicate that the jets were likely driven by magnetic force.
	\item The HXR source and hot plasma at the base of the jet were expected from current models. However, the HXR sources at the top of the jet were produced by energetic electrons that were accelerated very close to the top location, rather than electrons that were accelerated near the jet base but traveled to the top. This means that there was more than one reconnection and particle acceleration site in each event.	
\end{enumerate}	

Coronal jets are an important form of solar activity that involves particle acceleration, and they share similarities with larger eruptive events such as CMEs. HXRs can provide important constraints on hot plasma within a coronal jet, as well as unique diagnostics of energetic electron populations. To obtain the best constraints for jet models, observations should take advantage of state-of-the-art instruments in different wavebands, but only a few studies have included HXR observations to date. In future work, we would like to extend the method described in this paper to other coronal jets. Those jets could come from the jet database that will be generated by the citizen science project Solar Jet Hunter \footnote{https://www.zooniverse.org/projects/sophiemu/solar-jet-hunter} (which was launched through the Zooniverse platform in December 2021). We expect studies with more jet samples to further advance our understanding of particle acceleration in jets.

Furthermore, as shown in this study, HXR sources that are associated with jets could be found in the corona, and they could be faint in some events, thus not identified by current instruments. One solution is to develop direct focusing instruments, such as that demonstrated by the Focusing Optics X-ray Solar Imager (\textit{FOXSI}) sounding rocket experiment, which will provide better sensitivity and dynamic range for future HXR observations. 

\acknowledgments
This work is supported by NASA Heliophysics Guest Investigator grant 80NSSC20K0718. Y.Z. is also supported by the NASA FINESST program 80NSSC21K1387. N.K.P. acknowledges support from NASA's {\sdo}/AIA and HGI grant. We thank Samaiyah Farid for helpful discussions. We are also grateful to the {\sdo}/AIA, {\rhessi}, {\hinode}/XRT, and {\iris} teams for their open data policy. {\hinode} is a Japanese mission developed and launched by ISAS/JAXA, with NAOJ as domestic partner and NASA and STFC (UK) as international partners. It is operated by these agencies in co-operation with ESA and the NSC (Norway). {\iris} is a NASA small explorer mission developed and operated by LMSAL with mission operations executed at NASA Ames Research Center and major contributions to downlink communications funded by ESA and the Norwegian Space Centre.

\bibliography{jet}
\bibliographystyle{aasjournal}

\end{document}